\documentclass[twoside]{article}
\usepackage[a4paper]{geometry}
\usepackage[utf8]{inputenc} 
\usepackage[T1]{fontenc} 
\usepackage{RR}
\usepackage{hyperref}

\usepackage{cite}
\usepackage{amsmath,amssymb,amsfonts}
\usepackage{algorithmic}
\usepackage{graphicx}
\usepackage{textcomp}
\usepackage{todonotes}
\usepackage{url}
\usepackage{caption}
\usepackage{enumitem}
\usepackage{eurosym}
\usepackage[utf8]{inputenc}
\usepackage{xcolor}

\RRNo{9234}
\RRdate{December 2018}
\RRauthor{Mathieu Cunche~\thanks{Univ Lyon, INSA Lyon, Inria, CITI, F-69621 Villeurbanne, France} \and
Daniel Le Métayer~\thanks[sfn]{Univ Lyon, Inria, INSA Lyon, CITI, F-69621 Villeurbanne, France} \and
Victor Morel~\thanksref{sfn}}

\authorhead{Cunche \& Le Métayer \& Morel}

\RRtitle{Un cadre conceptuel pour l'information et le consentement dans l'Internet des Objets}

\RRetitle{A Generic Information and Consent Framework for the IoT (Extended Version)}

\titlehead{A Generic Information and Consent Framework for the IoT}

\RRresume{L'Internet des Objets (IoT) soulève des questions particulières en terme d'information et de consentement, ce qui rend l'implémentation du Règlement Européen sur la Protection des Données (RGPD) difficile dans ce contexte. Dans ce rapport, nous proposons un cadre conceptuel pour l'information et le consentement dans l'IoT à la fois protecteur pour les personnes concernées et les responsables de traitement. Nous présentons une description de haut niveau de ce cadre, illustrons sa généralité à travers différentes solutions techniques et études de cas, et ébauchons un prototype d'implémentation.
}
\RRabstract{The Internet of Things (IoT) raises specific issues in terms of information and consent, which makes the implementation of the General Data Protection Regulation (GDPR)  challenging in this context. In this report, we propose a generic framework for information and consent in the IoT which is protective both for data subjects and for data controllers. We present a high level description of the framework, illustrate its generality through several technical solutions and case studies, and sketch a prototype implementation.  }
\RRmotcle{vie privée \and IoT \and information \and consentement \and RGPD \and règlement}

\RRkeyword{privacy \and  IoT  \and information \and consent \and GDPR \and regulation.}
\RRprojets{Privatics}
\RCGrenoble 

\begin{document}
\makeRR   
\section{Introduction}

The development of the Internet of Things (IoT) raises specific privacy issues especially with respect to information and consent. Data subjects are generally unaware of the devices collecting data about them and do not have prior contact with the third parties operating them. Solutions such as stickers or wall signs are not effective information means since they remain unnoticed from most data subjects. As far as consent is concerned, data subjects do not have simple means to communicate with data controllers to express their privacy choices. Furthermore, the devices used to collect data in IoT environments have scarce resources; some of them do not have any user interface, are battery-operated or operate passively (collecting data without emitting any signal). 

The  General Data Protection Regulation (GDPR) \cite{gdpr-2016}  puts emphasis on the control of data subjects over their personal data. Its application  to the IoT is not obvious though. The Working Party 29\footnote{One of the roles of the Working Party 29, which is now replaced by the European Data Protection Board,  was to make recommendations on data protection and privacy in the European Community.}  (WP29) has  published  guidelines on transparency~\cite{wp29-transparency-2017} and consent~\cite{wp29-consent-2017} and an opinion on the development of the IoT~\cite{wp29-iot-2014}. Starting from these recommendations, we discuss in Section \ref{wp29} the specific challenges raised by the IoT in terms of transparency and consent and we propose in  Section \ref{framework} an abstract framework to address them. The framework is generic in the sense that it defines essential requirements that have to be met to ensure that information and consent are managed in a manner that is satisfactory  both for data subjects and for data controllers. In Section \ref{techniques}, we introduce several techniques to implement this framework in different situations, in particular through declaration registers and beacons. These techniques are illustrated with several challenging case studies in Section \ref{studies}. 
In Section \ref{implem}, we sketch a prototype implementation of these techniques.  
We discuss related work in Section \ref{related} and conclude with perspectives in Section \ref{conclusion}.  

\section{European Regulation and WP29 Recommendations} 
\label{wp29}

Over the last decades, the idea that individuals should have an effective control over their personal data has become a key part of the EU doctrine in the field of data protection. In many policy documents, control is advocated as an important tool for protecting privacy and achieving the empowerment of data subjects~\cite{lazaro-lemetayer-2015}. As an illustration, Recital 7 of the GDPR\cite{gdpr-2016} states that ``Natural persons should have control of their own personal data'' and the current draft of the new ePrivacy Regulation refers to the right for natural and legal persons to ``control electronic communications''. Control is not defined precisely in the GDPR but it is backed up by a number of provisions, including enhanced obligations for data controllers in terms of transparency and consent.

To facilitate the interpretation of the GDPR, the WP29 has  published two guidelines on transparency~\cite{wp29-transparency-2017} and consent~\cite{wp29-consent-2017}. The WP29 has also published two opinions that are relevant to this report, on the development of the IoT~\cite{wp29-iot-2014} and on the draft ePrivacy Regulation~\cite{wp29-eprivacy-2017} respectively.

As far as transparency is concerned, the GDPR defines the categories of information to be provided to data subjects: identity of the controller, purpose of the processing, categories of personal data concerned, recipients, etc. It also introduces some requirements on acceptable communication modes.
Recital 39  states  ``The principle of transparency requires that any information and communication relating to the processing of those personal data be easily accessible and easy to understand, and that clear and plain language be used''.  According to the WP29, ``the \textit{easily accessible} element means that the data subject should not have to seek out the information; it should be immediately apparent to them where this information can be accessed, for example by providing it directly to them, by linking them to it, by clearly signposting it or as an answer to a natural language question.'' The WP29 also suggests that IoT devices have a  QR code that can be scanned to display the transparency information. However, it is  questionable whether informing data subjects through QR codes or signposting is consistent with the idea, also put forward by the WP29, that ``the data subject must not have to take active steps to seek the information covered by these articles or to find it among other information''. We believe that the best way to protect the interests of both parties is to  make it possible for  data  controllers to communicate  all the required information  to data subjects in electronic form. Considering that IoT devices are by definition electronic objects collecting data from subjects, there is no  reason why   electronic means could not be used also to inform them. However, IoT raise specific challenges to this respect: there is a wide variety of devices, most of them have scarce resources, some of them, such as cameras, are passive or collect data on a permanent basis. Nevertheless, the systematic communication of the information to data subjects is not out of reach, neither from the technical point of view, nor from an economic standpoint, as discussed in the next sections.

Another key issue that has been discussed in many studies is the need to avoid user fatigue.  The WP29 states that ``The requirement that the provision of information to, and communication with, data subjects is done in a 'concise and transparent' manner means that data controllers should present the information communication efficiently and succinctly in order to avoid information fatigue''. The WP29 recommends in particular the use of  ``push'' and ``pull'' notices. As stated by the WP29, ``push notices involve the provision of 'just-in-time' transparency information notices while pull notices facilitate access to information by methods such as permission management, transparency dashboards and 'learn more' tutorials. These allow for a more user-centric transparency experience for the data subject.''

The GDPR also defines a number of conditions for the validity of consent: it should be freely given, specific, informed and unambiguous. These conditions also raise new challenges in the context of the IoT. For example, Recital 42 states that ``Consent should not be regarded as freely given if the data subject has no genuine or free choice or is unable to refuse or withdraw consent without detriment''. In the context of the IoT, this should entail that consent to physical tracking is not valid if the only alternative for  data subjects is to turn off their WiFi and thereby be deprived of useful services. To ensure the lack of ambiguity, consent should, according to the GDPR, ``be given by a clear affirmative act'', which should exclude the collection of identifiers such as MAC addresses for example, without any affirmative action from the user. These issues are all the more important for data controllers given that the GDPR requires that they must be able to demonstrate that a valid consent has been provided\footnote{This requirement holds only if the legal basis for the processing is consent. Data controllers can rely on other legal bases such as their legitimate interest or the use of the data for the execution of a contract.}.

The need to avoid user fatigue is also critical for consent. As stated by the WP29 in its guidelines on consent~\cite{wp29-consent-2017}, ``In the digital context, many services need personal data to function, hence, data subjects receive multiple consent requests that need answers through clicks and swipes every day. This may result in a certain degree of click fatigue: when encountered too many times, the actual warning effect of consent mechanisms is diminishing. This results in a situation where consent questions are no longer read. This is a particular risk to data subjects, as, typically, consent is asked for actions that are in principle unlawful without their consent. The GDPR places upon controllers the obligation to develop ways to tackle this issue.''

As far as the IoT is concerned, the WP29 advocates the design of new consent mechanisms, such as ``privacy proxies''\footnote{``In practice, today, it seems that sensor devices are usually designed neither to provide information by themselves nor to provide a valid mechanism for getting the individual’s consent. Yet, new ways of obtaining the user’s valid consent should be considered by IoT stakeholders, including by implementing consent mechanisms through the devices themselves. Specific examples, like privacy proxies and sticky policies, are mentioned later in this document.''~\cite{wp29-iot-2014}}, on the devices themselves. We agree that this is a key condition for the effective implementation of consent  and discuss this further in the next sections.

\section{Generic framework for information and consent} 
\label{framework}

The above analysis of the GDPR show that it raises two main categories of challenges in the context of IoT :
\begin{itemize}
\item Communication  between data subjects and data controllers: how to ensure that data subjects always receive the required information from the data controllers and that data controllers always receive valid consents from  data subjects?
\item Involvement of the data subjects in the process: can they read, analyse and understand all relevant information? Can they define their privacy choices carefully and thoughtfully? 
\end{itemize}

We derive from this analysis a set of technical requirements for implementations of information and consent  which are  protective both for data subjects (hereinafter ``DS'') and for data controllers (hereinafter ``DC''). We assume that DC deploy devices that can collect different types of personal data and/or communicate information to DS. For their part, DS may own several devices and at least one of them (typically a smartphone) can be used to consult the information provided by DC and to express their consent. We call this device the Gateway Device. We use the expression ``DS privacy policy'' to refer to the choices of the data subject regarding his personal data and ``DC privacy policy'' to refer to the privacy policy declared\footnote{A declaration can be seen as a commitment of the DC  to implement his DC  privacy policy but the actual enforcement of this policy is outside the scope of this paper.}  by a DC.  In order to meet the challenges identified in Section \ref{wp29}, a  consent management framework should provide the following facilities for, respectively, information and consent:

\vspace{0.2cm}

\textbf{Information:}
\begin{itemize}
\item The declaration by  DC of their devices, with all the necessary information, including their position, range, the type of data collected and the associated DC privacy policy.
\item The receipt of this information by the device of any DS about whom personal data can be collected (i.e. within the range of a DC device). 
\item The presentation of this information to the DS in forms and at times that should minimize information fatigue and maximize the likelihood that the DS will not miss any useful information.
\end{itemize}

\textbf{Consent: }
\begin{itemize}
\item Means for DS to express their consent in forms and at times that should minimize their fatigue and maximize the likelihood that they make appropriate decisions regarding the protection of their personal data.
\item The receipt of this consent by any DC able to collect data about DS and the guarantee that they will not collect the data (or will immediately delete it) if this consent is not consistent with their DC policy. 
\item The possibility for  DC to store the consents obtained from DS so  they can demonstrate GDPR compliance regarding consent, in particular that it has been emitted by the data subject on whom data is held. 
\end{itemize}

The interactions between a DC and a DS can be split into two parts: the interactions of the DS with his Gateway Device  (to be informed and to express his consent) and the communications between the DS Gateway Device and the DC devices. 

In the rest of this section, we define more precise requirements on the communications between DC devices and DS devices. 
We first define the operations considered here, which can be triggered either by a DC or by a DS (or their devices):

\begin{itemize}
\item $install(\delta, \lambda, \rho, \theta, \pi)$ is the deployment of a collecting device $\delta$ at position $\lambda$ with range $\rho$, collecting data of type $\theta$ with DC privacy policy $\pi$. The position and the range define the physical space in which the device can collect data of type $\theta$. The type can be for example MAC address, sound, or image. We assume without loss of generality that a device is associated with only one type\footnote{Multi-type devices can be considered as several devices at the same location.}.   
\item $declare(\delta, \lambda, \rho, \theta, \pi)$ is the declaration of a collecting device $\delta$ at position $\lambda$ with range $\rho$, collecting data of type $\theta$ with DC privacy policy $\pi$.
\item $collect(\delta, \sigma, \theta, \pi, \mu )$ is the collection by device $\delta$ of value $\mu$ of type $\theta$ from the DS device $\sigma$; the value is associated with the DS privacy policy $\pi$.
\item $move(\sigma, \lambda)$ means that the DS device $\sigma$ moves to position $\lambda$.
\item $define(\sigma, \theta, \pi, \mu)$ means that the DS privacy policy and the value of  data of type $\theta$ on device $\sigma$ are set to $\pi$ and $\mu$  respectively.
\item $pair(\sigma_1, \sigma_2)$ is the pairing of the DS device $ \sigma_1$ to the DS device $\sigma_2 $. Pairing is useful in this context to make it possible to define the privacy policy of a device $\sigma_1$ with scarce resources on another device $\sigma_2$ of the DS (his Gateway Device). 
\item $require(\sigma_1, \sigma_2, \delta, \theta, \pi, \mu) $ means that device $\sigma_1$ requires that $\delta$ updates the privacy policy and value of  data of type $\theta$ collected from $ \sigma_2$ to $\pi$ and $\mu$  respectively. The $require$ operation can be used by a DS for example to require the deletion of his data if he wants to withdraw his consent\footnote{A deletion request is a request with a policy $\pi $ having a retention delay equal to zero.}. Note that $\sigma_1$ is the device emitting the requirement and $\sigma_2$ is the device from which the data has been collected by the DC device $\delta $. In practice, it is often the case that $\sigma_1 = \sigma_2$ but we may also have $\sigma_1 \neq \sigma_2$ when the DS privacy policy is expressed and stored on a device (the Gateway Device represented by $\sigma_1$) and the  data is stored on another DS device (his watch for example, represented by $\sigma_2$).  
\end{itemize}

Some of the above operations ($install$, $move$) are physical while $collect$ describes the actual collection of the data. The key operations in terms of information and consent management are $declare$, $define$ and $require$. Different implementations of these operations are described in the next section.  Our goal at this stage is to provide a  high-level description of the framework.  For example, what is called a ``device''  here can be any source of personal data, such as a smart phone, a quantified-self device or even the subject himself for data of type voice or image. The identifier $\delta $ of a device can be implemented in many different ways (MAC address, plate number, etc.) as discussed in the next section and illustrated in Section \ref{studies}. Some of these devices cannot store their DS privacy policies, hence the need to distinguish $\sigma_1$ and $\sigma_2$ in the $require$ operation. Typically, the mobile phone of a DS can play the role of Gateway Device and therefore be used to define and store all his privacy policies (for all his devices).

We do not discuss in this paper the content of a privacy policy $\pi$. The definition of privacy policies and their semantics are studied in a companion paper~\cite{pilot2018}. In the present report, we make only two assumptions about these policies:
\begin{itemize}
\item DC policies contain all the information required by the GDPR (including the information about the DC, purpose of the collection, retention delay, etc.).
\item Policies have a well-defined semantics endowed with an implication relation $\succ$, with $\pi_1 \succ \pi_2$ meaning that policy $\pi_1$ is at least as strong as policy $\pi_2$.
\end{itemize}

In order to define the semantics, or precise meaning, of the above actions, we need to characterize the abstract state of the system. This state consists of the following functions:

 \begin{itemize}
\item $Config(\delta) = (\lambda, \rho, \theta, \pi)$ means that DC device $\delta$ is at position $\lambda$ with range $\rho$; it collects data of type $\theta$ with DC privacy policy $\pi$.   
\item $Declared_c(\delta) = (\lambda, \rho, \theta, \pi)$ means that  DC device $\delta$ has been declared with position $\lambda$ and range $\rho$, collecting data of type $\theta$ with DC privacy policy $\pi$. 
\item $Knows_s(\sigma, \delta) = (\lambda, \rho, \theta, \pi)$ means that  DS device $\sigma$ has been informed of the presence of a collecting device $\delta$ is at position $\lambda$ with range $\rho$, collecting data of type $\theta$ with DC privacy policy $\pi$. 
\item $Position(\sigma) = \lambda$ means that  DS device $\sigma$ is at position $\lambda$.
\item $Paired(\sigma_1) = \sigma_2$  means that  DS device $\sigma_1$ is paired to  DS device $\sigma_2$ which hosts its privacy policies. We assume by convention  $Paired(\sigma) = \sigma$  for devices $\sigma$ hosting their own policies. 
\item $Store_c(\delta, \sigma, \theta) = (\pi,\mu)$ means that   DC device $\delta$ stores the value $\mu$ of type $\theta$ with   DS privacy policy $\pi$  originating from device $\sigma$. 
\item $Store_s(\sigma, \theta) = (\pi,\mu)$ means that   DS device $\sigma$ stores the value $\mu$ of type $\theta$ with  DS privacy policy $\pi$.
\end{itemize}

In the same way as the operations, the state is defined at a very high level at this stage and it can be implemented in different ways, as discussed in the following section. Basically, the $Config$,  $Position$ and $ Paired$ functions describe the physical environment (position and features of the devices) whereas the other functions define the information stored by DC ($Declared_c$ and $Store_c$) and by DS ($Knows_s$ and $Store_s$). In practice, this information can be stored on the devices themselves, on a server, or a combination of both.

We can now use this abstract state to define the requirements on the operations. Each operation is associated with a precondition and a postcondition in the style of Hoare logic. The precondition is a property of the state that must be satisfied for the operation to occur and the postcondition is a property of the state that must be satisfied after the execution of the operation. Following the usual notation, we use primes to denote the state after the execution of the operation and we assume that the parts of the states that are not specified in the postcondition are not affected by the operation.  For example, the property  $Knows_s'(\sigma, \delta) = (\lambda, \rho, \theta, \pi)$ means that the state is unchanged except for $Knows_s(\sigma, \delta) $ which is set to $(\lambda, \rho, \theta, \pi) $. 

In the following, the notation $\star $ is used to denote any value and  function $\triangleright $ is defined as follows: if $ \alpha \neq  \bot $ then $ \alpha \triangleright \beta = \alpha $, otherwise $ \alpha \triangleright \beta = \beta$. Relation $Within(\lambda_s,\lambda,\rho)$ expresses the fact that position $\lambda_s$ is within the range of a device located at position $\lambda $  with range $ \rho$.
 
\vspace{0.3cm}

\textbf{Operation:} $install(\delta, \lambda, \rho, \theta, \pi)$

\textbf{Precondition:} $Declared_c(\delta) = (\lambda, \rho, \theta, \pi)$

\textbf{Postcondition:} $Config'(\delta) = (\lambda, \rho, \theta, \pi)$

\vspace{0.3cm}

\textbf{Operation:} $declare(\delta, \lambda, \rho, \theta, \pi)$ 

\textbf{Precondition:} $True$

\textbf{Postcondition:} $Declared_c'(\delta) = (\lambda, \rho, \theta, \pi)$ $ \wedge $

 $ \forall \lambda_s, \sigma, ((Within(\lambda_s,\lambda,\rho)$ $  \wedge$ $ Position(\sigma) = \lambda_s)  \Rightarrow$
 
$Knows_s'(\sigma, \delta) = (\lambda, \rho, \theta, \pi)) $  

\vspace{0.3cm}

\textbf{Operation:} $collect(\delta, \sigma, \theta, \pi, \mu )$ 

\textbf{Precondition:} $\exists \lambda, \rho, \pi_c, \lambda_s, \pi_1, \mu_1,$

$ Config(\delta) = (\lambda, \rho, \theta, \pi_c)$ $ \wedge$ $
Store_c(\delta, \sigma, \theta) = (\pi_1,\mu_1) $ $\wedge $

$ Position(\sigma) = \lambda_s 
 \wedge Within(\lambda_s,\lambda,\rho)$ $ \wedge$ $ Store_s(\sigma, \theta) = (\pi,\mu) $ $ \wedge$ $ \pi \triangleright \pi_1 \neq \bot $

\textbf{Postcondition:} $ (\pi_c \succ (\pi \triangleright \pi_1)) \Rightarrow Store_c'(\delta, \sigma, \theta) = ((\pi \triangleright \pi_1),\mu) $ $\wedge$ $  (\pi_c \not\succ (\pi \triangleright \pi_1)) \Rightarrow Store_c'(\delta, \sigma, \theta) = (\bot ,\bot) $

\vspace{0.3cm}

\textbf{Operation:} $move(\sigma, \lambda)$ 

\textbf{Precondition:} $True$

\textbf{Postcondition:} $Position'(\sigma) = \lambda$  $\wedge $ 

$\forall \delta, \lambda_c, \rho, \theta, \pi,
 (Declared_c(\delta) = $ $(\lambda_c, \rho, \theta, \pi) $ $\wedge$  
 
 $Within(\lambda,\lambda_c,\rho)$) $ \Rightarrow $  $Knows_s'(\sigma, \delta)$ $ = (\lambda_c, \rho, \theta, \pi)$

\vspace{0.3cm}

\textbf{Operation:} $define(\sigma, \theta, \pi, \mu)$

\textbf{Precondition:} $True$ 

\textbf{Postcondition:} $Store_s'(\sigma, \theta)$ = $(\pi\triangleright\pi_1,\mu\triangleright\mu_1)$

with $ Store_s(\sigma, \theta) = (\pi_1,\mu_1)$

\vspace{0.3cm}

\textbf{Operation:} $pair(\sigma_1, \sigma_2)$ 

\textbf{Precondition:} $True$

\textbf{Postcondition:} $Paired'(\sigma_1) = \sigma_2$ 

\vspace{0.3cm}

\textbf{Operation:} $require(\sigma_1, \sigma_2, \delta, \theta, \pi, \mu) $

\textbf{Precondition:} $\exists \lambda, \rho, \pi_c, \pi_1, \mu_1, \lambda_s, $

$Config(\delta) = (\lambda, \rho, \theta, \pi_c)$ $ \wedge $ $Store_c(\delta, \sigma_2, \theta)$ $ = $ $(\pi_1,\mu_1) \wedge Position(\sigma_1) = \lambda_s $ $\wedge $ $ Within(\lambda_s,\lambda,\rho) $ $\wedge $ $ Store_s(\sigma_1, \theta) = (\pi,\star) $
$\wedge $ $ Store_s(\sigma_2, \theta) = (\star,\mu) $ $\wedge $ $Paired(\sigma_2) = \sigma_1$

\textbf{Postcondition:} $ Store_c'(\delta, \sigma_2, \theta) $ $ = (\pi\triangleright\pi_1,\mu\triangleright\mu_1)$

The preconditions for  $declare$, $move$ and $define$ are equal to $True$ because a DC can always declare a new device and a DS can always decide to move or to modify his personal data or privacy policies. The precondition for $install$ captures the requirement that all installed devices must be declared. The precondition for $collect$ expresses the fact that the DS device $\sigma $ must be within the range of the collecting device $\delta $. Note that some of the values in the state can be undefined. For example, $ \pi_1 = \bot$ if the DC does not have any policy for data of type $\theta $ originating from $\sigma $. The intuition for the precondition of $require$ is similar except that two paired DS devices $\sigma_1$ and $\sigma_2$ are involved\footnote{As discussed above, we have $\sigma_1 = \sigma_2$ for devices hosting their privacy policies.} : $\sigma_1$ is the device hosting the policy ($ Store_s(\sigma_1, \theta) = (\pi,\star) $) and  $\sigma_2 $ the device hosting the data ($Store_s(\sigma_2, \theta) = (\star,\mu)$). 

The postconditions of $declare$ and $move$ capture the requirement that any DS device $\sigma $ within the range of a DC device $\delta $ must receive the declaration made by the DC (which is expressed by $Knows_s'(\sigma, \delta) = (\lambda, \rho, \theta, \pi))$ for $declare$). Function $\triangleright $ is used to take into account the fact that certain parameters may be undefined. For example, if data is collected (through $collect$) without a privacy policy, we have $\pi = \bot $  and  $\pi \triangleright \pi_1 = \pi_1 $. This means that the policy associated with the data is $ \pi_1$, which has already been collected by the DC ($ \pi \triangleright \pi_1 \neq \bot $, from the precondition, therefore $\pi_1 \neq \bot $). The postcondition of $collect$ also expresses the requirement that a DC cannot store any data associated with a DS privacy policy that is not met by his own DC policy ($  (\pi_c \not\succ (\pi \triangleright \pi_1)) \Rightarrow Store_c'(\delta, \sigma, \theta) = (\bot ,\bot)$). 

From the above requirements, we can prove key properties such as the following:
\begin{itemize}
\item No collection of data can take place if the DS has not previously received  the required information from the DC.
\item As long as DS do not change their DS privacy policies, any data collected by a DC is associated (in the state of the DC) with the privacy policy of the DS.
\item If a DS  changes his privacy policies, any data collected by a DC is associated (in the state of the DC) with the last privacy policy received from the DS.
\item No data is collected from a DS and stored by a DC if the current DS privacy policy is not met by the DC policy. 
\end{itemize}

\section{Technical options} 
\label{techniques}

The requirements introduced in the previous section are very high-level and can be implemented in different ways. In this section, we present technical options to implement them depending on the context and the capacities of the DC and DS devices. These techniques are illustrated through several case studies in  Section \ref{studies}. We first describe the implementation of communications between DC devices and DS devices, which corresponds to the $declare$, $collect$ and $require$ operations. Section \ref{direct} considers direct communications between devices while Section \ref{indirect} focuses on indirect communications (through registries). For each solution, we discuss its compliance with the above requirements and its feasibility in terms of cost and deployment effort. Then, we suggest ways to allow  DS to interact with their devices through a Personal Data Custodian in Section \ref{assistant}. These interactions concern in particular the implementation of the $define$ operation (expression of consent). They  also  contribute  to the information of the DS (operation $declare$). 

\subsection{Direct communications}
\label{direct}

\begin{figure}[!ht]
\centering
\includegraphics[scale=0.5]{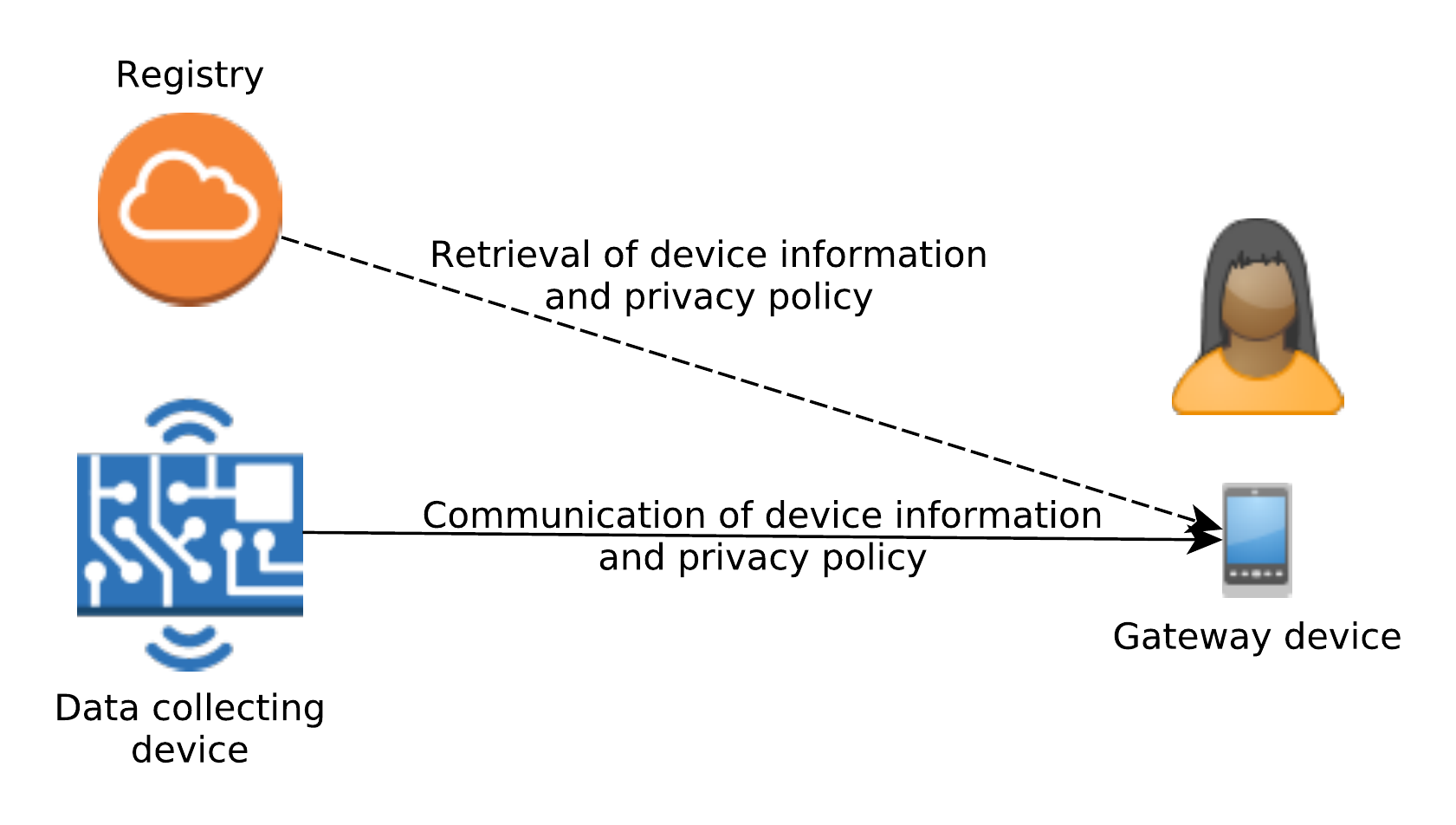}
\caption{Direct and indirect declarations}
\label{fig:transparency}
\end{figure}

A first option to implement information and consent is through direct communications between DC devices and the DS Gateway Device. 
In this option (hereinafter ``direct communication''), DC devices use a direct communication channel to advertise their presence and communicate all the parameters of the $declare$ operation (position, range, type of data collected and DC privacy policy) within their area of operation. The same communication channel can be used by the DS to transmit his potential consent to the DC.

Direct communication can typically be implemented using medium and short range wireless communications technologies such as  Bluetooth or Wi-Fi which are now common place and are embedded in many devices (e.g. smartphones).  In addition, their range (typically several meters to tenths of meters) matches the scale of the area of operation of IoT systems and their protocol can be leveraged to carry the information required for declaration and consent. In this section, we focus on the BLE technology, but other wireless technologies could be used in a similar way.

Bluetooth Low Energy (BLE) features a discovery mechanism that allows the detection and identification of devices as well as the transmission of small amounts of data. This mechanism can be used to implement direct communications between the DC and the DS Gateway Device. In the following we refer to the element implementing this mechanism on the DC side as the \emph{BLE Privacy beacon}.

In Bluetooth Low Energy (BLE), device discovery is implemented using  \emph{Advertisement Packets}~\cite[Part C, sec. 11]{bluetooth_sig_specification_2016} that are broadcast at regular intervals and  can be received by any BLE device in range. Those packets can be configured to carry data necessary for the declaration of DC devices (parameters of the \emph{declare} operation). A DS Gateway Device in the range of the DC \emph{BLE Privacy beacon} will thus be able to passively retrieve the declaration data by collecting the advertising packets and extracting the relevant information.

Another feature offered by BLE is the Attribute Protocol (ATT)~\cite[Part A, sec. 6.4]{bluetooth_sig_specification_2016} that allows the exposure of services and the transmission of small amounts of information through a lightweight connection.  With ATT, a BLE device can have a set of \emph{profiles}, each of them exposing a list of \emph{attributes} that can support  read and write operations. 
This feature can be leveraged to implement the communication of consent: the DS Gateway Device connects to the \emph{BLE Privacy Beacon}  (this is a lightweight process) and send the consent data (parameters of the \emph{define} operation) using the ATT protocol.


Direct communications have several benefits: first, they do not require an Internet connectivity. Also, the locality of the  communications reduces the risk of further tracking by a remote entity. From the point of view of the DS, the information part is collected passively by collecting the data transmitted by the \emph{BLE Privacy Beacon}; this means that, in order to be informed,  the DS does not  expose his presence. 
They also raise several challenges. First,
all devices should be able to  declare themselves. Tracking systems involving passive devices thus need to be enhanced (for example with a beacon) to enable these declarations. Also, the communication protocol should support the communication of the parameters of the $declare$ and $define$ operations. 
In addition, the communication range of the declaration should match the operational range of the data collection by the device. 
Finally, all the above features should be possible at reasonable cost and without disrupting existing services.



\subsection{Indirect communications}
\label{indirect}

Another option to implement information and consent is to use a registry (hereinafter ``indirect communication'').  Registries can be used both by DC  (implementation of the \emph{declare} operation) and by DS (implementation of the \emph{define} operation).  A DC registry is a database freely accessible through the Internet, storing all relevant information about DC devices, including the parameters of the $declare$ operation. The DC registry declaring a DC device $\sigma_c$ must be accessible to any DS device $\sigma_s$ (or the paired Gateway Device) before it enters the range of this DC device  (i.e. when $ Within(\lambda_s,\lambda_c,\rho_c)$ if $\lambda_s$ is the location of $\sigma_s$ and $\lambda_c$ and $\rho_c$  are respectively the location and range of $\sigma_c$), for example  via a web site or through an application. They must provide the required information in  machine-readable format, for instance a structured format such as JSON provided through an API. They should also include a human-readable version that can be consulted directly by DS. 

Indirect communications through DC registries have several advantages compared to direct communications: (1) they enable the visualization of DC policies regardless of the location of DS, which means that DS can be informed about the collection of data before visiting an area, (2) they provide a flexible management approach for DC policies — they do not require a specific infrastructure or particular capabilities of the devices except for an Internet connection\footnote{Therefore, they can be well-suited to passive devices such as cameras.}, and (3) they do not require the implementation of interactions with DS when a legal ground other than consent is considered.

However, their implementation raises several challenges: (1)
DS devices must always be  aware of all surrounding devices; therefore, inconspicuous or difficult to access registries are not acceptable.
(2) registries must be properly managed, up-to-date and accurate in order to meet the requirements defined in the previous section. Managing a registry can be achieved in different ways: it can be centralized or distributed, and contributions can be restricted to authenticated parties. 

In the same spirit, DS registries can be used by DC to retrieve the consents provided by DS. DC should be able to prove that the retrieved consents have effectively been provided by the right DS. This proof of identity could be implemented by using an authentication mechanism (\textit{e.g.} with tokens).

\subsection{Personal Data Custodian}
\label{assistant}

The interactions described in the previous sections involve a variety of devices. However, the ultimate recipients of DC policies and original sources of DS policies are the DS themselves.  In order to describe the interactions between DS and their devices, we assume that a software tool, called the Personal Data Custodian (hereinafter ``PDC''), is installed on DS Gateway Devices. The roles of the PDC  are  the following:

\begin{itemize}
\item Interact with the DS to allow him to consult the information received from DC devices (pursuant to the $declare$ operation). 
\item Interact with the DS to allow him to express his privacy choices (implementation of the $define$ operation). 
\item Interact with DC devices to communicate personal data with their DS policies  (implementation of the $collect$ operation) or to reject collection requests from DC devices when the associated DC privacy policy does not comply with the DS privacy policy.
\end{itemize}

As discussed in Section \ref{wp29}, consent is valid only if it is freely given, specific, informed and unambiguous. Each of these conditions brings forward strong requirements on the PDC and the language used to express privacy policies:

\begin{itemize}
\item \textit{Consent must be freely given:} any personal data and privacy policy communicated by the PDC should reflect the genuine choices of the DS.  
\item \textit{Consent must be specific:} the privacy policy language must be rich enough to allow DS to express granular choices, for example about types of data, data controllers or authorized purposes.  
\item \textit{Consent must be informed:} the PDC must not disclose personal data to a DC device that has not communicated its  privacy policy.
\item \textit{Consent must be unambiguous:} in order to avoid any ambiguity, the privacy policy language should be endowed with a formal semantics and the interface used to interact with the DS should not give rise to any misunderstanding.
\end{itemize}

A privacy policy language meeting these requirements is described in a companion paper~\cite{pilot2018}. In the present paper, we focus on the PDC itself and its interactions with the DS. The main challenge is to find the appropriate level of automation and type of interaction to meet the GDPR requirements while avoiding information fatigue. If the level of automation is low and interactions too frequent, consents may apparently meet all the GDPR requirements but in fact result from routine, mechanical, acceptance from DS, as already observed on the web. If the level of automation is  high, the reason may be that privacy policy rules are defined in a very coarse way (for example, ``always accept the disclosure of my \textit{MAC-address}'') which would not  meet the GDPR requirements.
Careful PDC design choices can help resolve this tension. For example, DS should be able to express positive rules (conditions in which they agree to communicate  a type of personal data) and negative rules (conditions in which they refuse to communicate  a type of personal data). The PDC would then interact with the DS only in situations for which it has not received any instruction (for example when an unknown category of DC device requests personal data). If the privacy language makes it possible to define choices at different levels of granularity, the DS can exploit this possibility to express generic consents (such as ``always accept the disclosure of my \textit{MAC-address} to a DC of category \textit{Museum} for the purpose \textit{Counting-visitors} only''. The PDC should then be able to instantiate this generic consent to a specific consent for a given museum visited by the DS. 

DS should also have the possibility to give their consent or dissent once for a particular occasion, but to be requested for future occurrences. A PDC should then propose to express temporary rules, regardless of the privacy language used. 

Different choices can also be made regarding the user interface provided to the DS. One option is to rely on a dashboard on which the DS can consult the DC declarations and  express his consent  through drop-down menus. Another option is to use a natural language version of the privacy policy language that can  be easily understood by the DS as suggested in~\cite{lemetayer-simple-2008,pilot2018}. This type of language can also be used to display DC declarations in a readable form.

\section{Case studies} 
\label{studies}

In order to illustrate the versatility of our framework and the technical options introduced  in Section \ref{techniques}, we present their application to a challenging vehicle tracking case study in Section  \ref{cs1}. We discuss more briefly other applications in Section  \ref{cs2}. We emphasize that all the solutions proposed here apply without any registration phase or assumption about prior  contacts between DS and DC.

\subsection{Vehicle tracking}
\label{cs1}

Vehicles may be subject to passive data collection by systems that detect and record their presence. One of the main technologies allowing this data collection is Automatic Number Plate Recognition (ANPR) \cite{du_automatic_2013} based on images captured by CCTV cameras. This technology has found many applications (e.g. law enforcement, road monitoring, parking billing and targeted advertisement). However, it has also  triggered serious privacy concerns as it involves the collection of large amounts of mobility data~\cite{noauthor_automated_2017}. 

Depending on the purpose of the ANPR system, the requirements regarding information and consent may vary. For instance, when used for billing purpose,  consent is not required because contract can be considered as the legal ground for data processing. However, consent would be required for ANPR systems collecting data for additional purposes such as profiling users to provide personalized services\footnote{For example, suggestions about parking places or routes.}. Even though they can be implemented independently, we consider in the following that  both information and consent are required. 

Let us consider a driving area (road section or parking lot) in which an ANPR system is operating, i.e.  all the vehicles in this area can have their plate number recorded and associated with other data (time, location, etc.). Let us further assume that the DC in charge of the ANPR system has no prior link with the DS. This means that, before the DS enters the area, the DC and the DS have not been able to communicate and that any exchange of information must be done on the spot. For what concerns the information of the DS, written signs could be displayed on the side of the road but they might remain unnoticed and would not convey the level of information required in this context.  In this challenging situation, the technical solutions presented in Section~\ref{techniques} can be used to implement mechanisms for seamless information and consent.

\begin{figure}
\centering
\includegraphics[width=10cm]{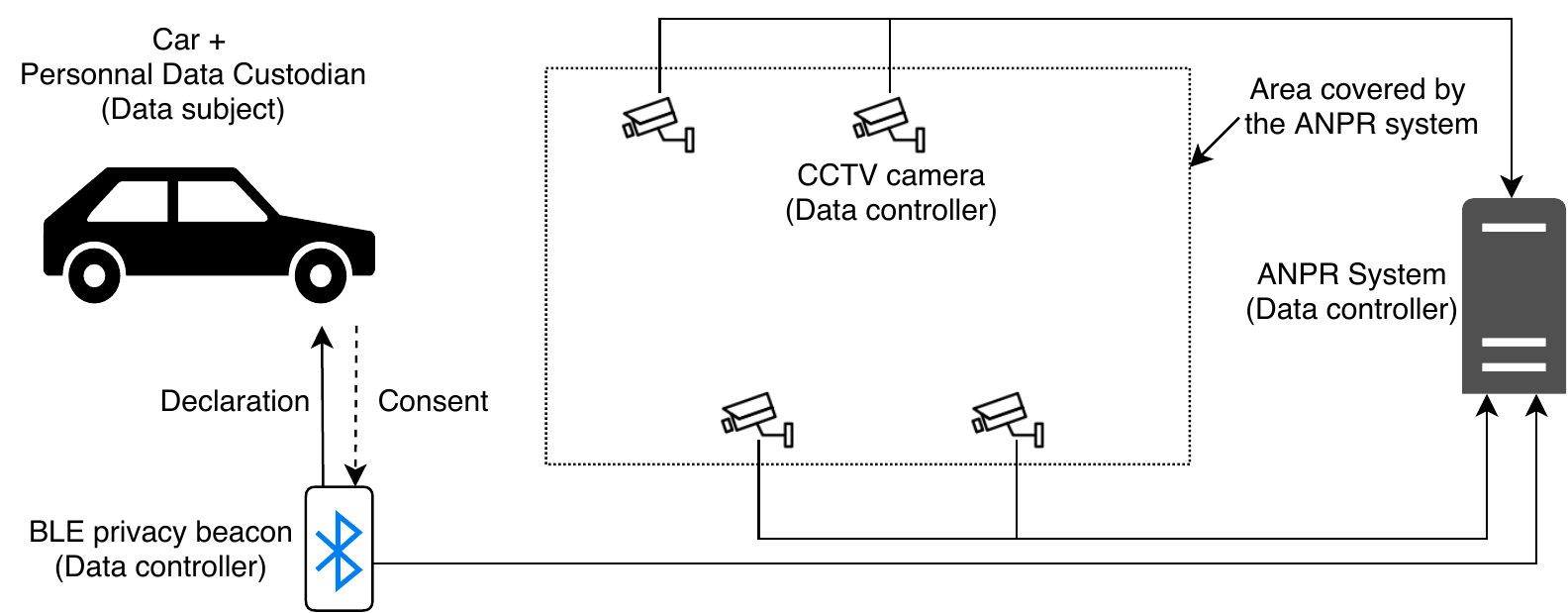}
\caption{Illustration of an ANPR system augmented with a BLE-based direction communication mechanism. If the privacy policy received from the DC complies with the DS privacy policy, the PDC of the DS automatically sends the consent (with the plate number) through the Bluetooth direct communication channel.}
\label{fig:ANPR_direct}
\end{figure}

The first option to address this case is to use direct communications via \emph{BLE Privacy Beacons} as presented in Section~\ref{direct}. Such beacons can be deployed in the ANPR area and around it in order to inform DS as soon as they enter the area. Furthermore, if the DS has configured his PDC with the relevant identifier (in this case the vehicle plate number), the PDC will also be able to use direct communications with the beacon in order to transmit a potential consent to the DC. The scenario is the following (see Figure~\ref{fig:ANPR_direct}): when entering the location, the PDC of the DS will automatically detect the \emph{BLE Privacy Beacon} and retrieve the privacy policy of the ANPR system  thanks to the direct communication mechanism. If this DC privacy policy complies with the DS privacy policy, the PDC of the DS automatically sends the consent through the Bluetooth direct communication channel. This consent contains  the plate number, which is the identifier of the DS in the context of this data collection. Once this consent is received and securely stored, the ANPR system is allowed to collect data on the vehicle identified by this plate number. By default, the ANPR system  discards any data for which it has not obtained consent. 

Another option to deal with APNR systems is to rely on indirect communications as described in section~\ref{indirect}. In this case, the DS is informed of the presence of the ANPR system through a DC registry: the DS Gateway Device regularly sends requests with its current localization to a dedicated server in order to get the list of nearby systems and the associated DC privacy policies. If the DC privacy policy complies with the DS privacy policy, the PDC sends the DS consent automatically to the dedicated online DS registry.

\subsection{Other applications}
\label{cs2}

\underline{Wi-Fi/Bluetooth tracking in shopping mall}: 
System monitoring the visitors of a shopping mall through Wi-Fi or Bluetooth are becoming commonplace. These systems passively collect the identifiers (MAC addresses) found in the messages broadcast by portable devices~\cite{demir:hal-00983363}. Any person entering the mall should be informed and should be able to easily provide his consent. This is particularly challenging in this kind of venue where there is generally no existing link between visitors and the managing entity; as a a result visitors are currently informed of those tracking system via posters and consent requirement is simply ignored. This use case is challenging but it can be addressed in our framework either in the direct mode or in the indirect mode.  In both cases, DS can be informed that Wi-Fi/Bluetooth data collection is taking place and can in turn send their consent, including their radio identifiers, through their PDC (if their own privacy policy allows for this collection).


\underline{Smart Meeting-room}:
Smart meeting-rooms are equipped with a variety of sensors and actuators in order to provide services to their hosts~\cite{waibel_smart:2003}. For instance, video cameras can record the position and movement of guests in the room, microphones can record the ambient sound to infer the nature of the discussion. We assume that this data can be collected only if all people in the room have provided their consent. The challenge here is to ensure that consent is retrieved from all guests, and that this consent is communicated with as little effort as possible. In this context, our approach could be applied as follows. Individual information and potential consents would be transmitted through direct or indirect communications as described in Section~\ref{techniques}. In parallel, a counting sensor is used to maintain an accurate number of guests in the room. The collection and processing of data within the room do not take place unless the DC has received a consent from as many people as there are in the room. In addition, if someone leaves or enters the room, the counting sensor is able to take the evolution into account and check the condition again.

We note that applying the indirect communication approach would yield in a solution similar to the one of Sadeh et. al \cite{sadeh_privacy_2017}. However, the direct communication approach would be significantly differ as all the exchanges of information will happen in the vicinity of the room, removing the need for network connectivity and the exposure of information to a third party.

\section{Prototype implementation}
\label{implem}

In this section, we briefly describe our prototype implementation of the direct communication solution presented in Section~\ref{direct}. We use a BLE Privacy beacon combined with a mobile application running on an Android phone implementing the PDC presented in Section {\ref{assistant}. The BLE Privacy beacon is based on a low cost (less than \$6) hardware (\emph{Espressif ESP32}\footnote{\url{https://www.espressif.com/en/products/hardware/esp32/overview}}) that implement the information and consent mechanisms (code of this BLE privacy beacon  is available online~\footnote{\url{https://github.com/cunchem/BLE_Privacy_Beacon.git}}.). 

The mobile application acts as a PDC and enables the definition of DS privacy policies in a user-friendly manner. 
So far, only the positive rules mentioned in Section \ref{assistant} are implemented. DS can add, update and delete rules through a scroll-down menu. 
The applications implements the direct communications to exchange privacy policies and consent with BLE Privacy beacon. 
Upon reception, the DC policy is compared to the current DS policy, and the PDC issues a consent message in case of compliance.

Our BLE device has been tested with 46 bytes long DC privacy policies. BLE devices are detected more than 10 meters away in indoor environment comprising load-bearing walls, and DC privacy policies are retrieved in less than 1 second after the PDC comes in range. The PDC is able to compare privacy policies and sends consent if the DC privacy policy complies with the DS privacy policy. 


\section{Related Work} 
\label{related}

In ~\cite{pappachan_towards_2017} the authors introduced the concept of \emph{IoT Resource Registries (IRRs)} that broadcast data  collection  policies and  sharing  practices  of  the local  IoT systems. Our direct communication mode has some similarities with this proposal. However, in contrast with~\cite{pappachan_towards_2017}, we provide technical solutions for its implementation and demonstrate an actual prototype. Furthermore, our approach is not limited to information since it also includes consent.

The Privacy Assistant project led by the Carnegie-Mellon University (CMU) is an example of use of registries to declare and retrieve privacy policies of IoT devices \cite{Das2018}. A prototype has been deployed on the CMU campus, where DS are able to locate cameras. Combined with an assistant on a mobile phone, subjects are warned about personal data collection in their vicinity. 
Our contributions with respect to this work are threefold: first, we present in Section \ref{framework} a generic framework that can be instantiated in different ways as shown in Section \ref{techniques}. The registries proposed in \cite{Das2018} represent one of the possible technical options. Also, we emphasize the control of DS over the disclosure of their personal data through their Gateway Devices (rather than through external privacy enforcement points). Last but not least, our first motivation is the implementation of the GDPR requirements in the context of IoT. As argued in Section \ref{framework}, the design choices of our framework are driven by this objective, especially the need to ensure that the criteria for valid consent are met.

Another example of registry dedicated to the declaration privacy choices is the Smart Places\footnote{\url{https://smart-places.org/}} service proposed by the Future of Privacy Forum (FPF).  With this service a data subject can provide his Wi-Fi or Bluetooth MAC address in order to opt-out from tracking by participating companies.



Previous work has also been done on privacy assistants\cite{Das2018}, including the PawS architecture\cite{Marc2002} and IoTA\cite{Das2018,Das2017}. In contrast with previous work in this area, our framework does not make any assumption about policy enforcement points and prior contact (e.g. through registration) between DS and DC: beacons announce themselves for direct communications, and registries are automatically retrieved by the PDC for indirect communications. Our framework enables local communications: DS policies does not have to be disclosed. In addition, our framework and privacy policies rely on a formal semantics which makes it possible to reason about privacy policies and provide further guidance to DS (for example abut the risks related to their policies)\cite{pilot2018}. 

In a nutshell, the main contributions of this paper with respect to previous work are the following: 
\begin{itemize}
\item Our framework is generic, including both direct and indirect communication modes, for both information and consent.
\item It is able to deal with the situation, which is common in the IoT, where DS do not have any prior contact with DC.
\item It relies on a formal semantics which makes it possible to avoid ambiguities and to provide formal guarantees. 
\item It has been devised to meet the requirements for information and consent in the GDPR and to enhance local control of the DS over the collection of their personal data. 
\end{itemize}

\section{Conclusion} 
\label{conclusion}

Beyond GDPR compliance, we believe that the adoption of the measures suggested in this paper would contribute to reduce the imbalance of powers between DC and DS without introducing prohibitive costs or unacceptable constraints for DC. The effectiveness of these solutions also depends on organizational and regulatory measures. For example, DC deploying or using IoT devices must have the legal obligation to declare their devices (with the required information) using electronic means~\footnote{The use of electronic means is not required by the GDPR and, so far, the WP29 seems to consider signposts as an acceptable information means.}. These solutions  also require a standardization effort (e.g. about the declaration protocol and the privacy policy language).

On the technical side, further work is required to improve the user-friendliness of the interface of our PDC and also to make it easier for DC to declare their devices. The fact that our framework and privacy policy languages are endowed with formal semantics also paves the way for richer user interfaces. For example,  we suggest in a companion paper~\cite{pilot2018} the verification of properties based on different risk assumptions. This facility could be useful to enhance DS awareness and to allow them to make better informed decisions.

The proposals made in this paper are also very relevant to the ongoing discussions about the future ePrivacy Regulation~\cite{eprivacy-2017}. As stated by the WP29~\cite{wp29-eprivacy-2017}, the current draft ``gives the impression that organisations may collect information emitted by terminal equipment to track the physical movements of individuals (such as Wi-Fi-tracking or Bluetooth-tracking) without the consent of the individual concerned.''The text is still evolving but this would be all the more unacceptable that, as discussed in this paper, solutions can be developed to make information and consent more effective, without introducing excessive constraints neither for data controllers nor for data subjects.

\section*{Acknowledgments}
This work has been partially funded by the CHIST-ERA project UPRISE-IoT, and by the INSA Lyon - SPIE ICS chair on the Internet of Things.

\bibliographystyle{plain}
\bibliography{biblio}
\end{document}